\documentstyle[psfig]{l-aa}
\begin{document}
   \thesaurus{12         
	      (11.05.2;  
               11.17.3;  
               12.03.2)} 

   \title{The evolution rate of quasars at various redshifts} 

   \author{Evanthia Hatziminaoglou, Ludovic Van Waerbeke, Guy Mathez}

   \offprints{E. Hatziminaoglou, eva@obs-mip.fr}

   \institute{ Observatoire Midi-Pyr\'en\'ees, Laboratoire 
d'Astrophysique, UMR 5572, 14 Avenue E. Belin, F-31400 Toulouse, France }

\date{Received, , Accepted,}


\maketitle

\markboth{E. Hatziminaoglou et al.: The evolution rate of quasars at
various redshifts}{}

\begin{abstract}
The evolution of optically selected quasars is usually supposed 
to be well described by a single constant evolution parameter, 
either $k_L$ or $k_D$, depending whether we refer to luminosity or 
density evolution. In this paper we present a study 
of the variations of the evolution parameters with redshift, for 
different cosmological models, in order to probe the differential 
evolution with redshift. Two different quasar samples have been analyzed, 
the AAT Boyle's et al. and the LBQS catalogues. Basically, these samples 
are divided in redshift intervals and in each of them $k_L$ and $k_D$ 
are estimated by forcing that $\langle V/V_{max}\rangle=0.5$. The 
dependence with respect 
to the cosmological parameters is small. Both AAT and LBQS show roughly the
same tendencies. LBQS, however, shows strong fluctuations, whose origin is
not statistical but rather due to the selection criteria. A discussion 
on selection techniques, biases and 
binning effects explains the differences between these
results. We finally conclude that the evolution parameter is almost constant
in the redshift range $0.7 \leq z \leq 1.7$, at least within $2\sigma$,
while it decreases slightly afterwards. Results
depend on the binning chosen (but not in a very significant way).
The method has been tested with 
Monte-Carlo simulated catalogues in order to give a better understanding of
the results coming from the real catalogues.
A correlation between $k_L$ ($k_D$) and $\langle V/V_{max}\rangle$ is
also derived and is used for the calculation of the error bars on the evolution
parameter. 

\end{abstract}

\section{Introduction}
\label{intro}
Since the first application of the $V/V_{max}$ test by Schmidt in 1968,
the luminosity function of quasars is known to undergo a strong 
evolution, in the sense that the density of the most luminous quasars
was far higher in the past. This was firstly interpreted in terms of either
Pure Density (PDE - Schmidt, 1968) or Pure Luminosity Evolution (PLE -
Mathez, 1976), and phenomenologically modeled with a single free parameter 
law. More complex models began to appear a few years later, such as a
Luminosity Dependent Density Evolution (LDDE), proposed by Schmidt and
Green in 1983. However, there wasn't any privileged model until the late
80's, when Boyle et al. (1988) favored PLE as a best fit of the observed 
luminosity function of a UVX sample. 
A similar recent result is found for X-Ray selected samples of
quasars (Page et al. 1997; \cite{Ja}; \cite{Ba97}).
The currently observed evolution of quasars appears however to be more complex 
nowadays, since the first indication of a reversing evolutionary
trend around a redshift in the range [2.5,3] was observed (Shaver, 1994;
Schmidt, Schneider and Gunn 1995; Warren, Hewett and Osmer 1994; Pei, 1995).
A steepening of the luminosity function towards high redshifts
is advocated by Goldschmidt and Miller (1997). 

Meanwhile, the evolution of the cosmic star formation density was found to
be strikingly similar to that shown by Shaver for QSOs, although the maximum 
of the star formation density is attained at a lower redshift (Madau, 1996). 
In spite of this difference, the similarity of the variations of
QSO luminosity density and of the field galaxy star formation has been
interpreted as a clue that both phenomena could be closely linked (Boyle
and Terlevitch, 1998; Silk and Rees 1998). Furthermore, the similarity
is even more striking (both curves have a maximum around the redshift z=2.5)
after applying the necessary  correction for dust
extinction in the density of high redshift galaxies (\cite{Sa98}). 
This correction may be as high as a
factor of ten since these galaxies are observed in their rest-frame
ultraviolet.

In the same time, complex models of this evolution begin to appear:

$\bullet $  At high redshift, the growth of Dark Matter halos according to Press-Schechter 
formalism and the parallel Eddington-limited growth of accreting Massive Black 
Holes (MBH) (\cite{HL}) are likely to
induce a decrease of density with increasing
redshift ('negative DE') (Haiman and Loeb, 1997; Cavaliere and Vittorini,
1998; Krivitsky and Kontorovich, 1998; Novosyadlyj and Chornij, 1997). 

$\bullet $  Around a redshift z=2.5, there could be two phenomena: 
a transition from
high to low efficiency in advection-dominated flows, followed by the
decline of accretion rate, giving luminosity proportional to $(1+z)^k$ 
with $k$ slightly variable around 3 (\cite{Yi}). Alternatively, growing
galaxies could assemble in groups of typically $5 \; 10^{12} \, M_{\odot}$, 
where tidal effects refuel the MBH at lower and lower rate,
translating into 'positive LE' (\cite{CV}).

$\bullet $  At intermediate redshift, viscous instabilities induce long term,
high amplitude variations of the accretion rate. The fraction 
of time spent at each luminosity level, convolved with the mass distribution
gives the luminosity function and its evolution (\cite{SE}). Galaxy 
collisions provide a mechanism which fuels galactic nuclei with gas in
dense environments, giving raise to quasars in low luminosity galaxies
(\cite{LKM}).
 
$\bullet $  In addition, gravitational lensing mimics luminosity evolution in
flux-limited samples, amplifying the more distant quasars, but whose
detailed effect is not known, in particular the induced bias selection
effects in magnitude limited quasar samples.

Van Waebeke et al. (1996) define a new test, the $V/V_{max}$ statistics, and 
apply it to the AAT quasar sample (Boyle et al., 1990). 
They show that, contrary to the usual $\langle V/V_{max}\rangle $ test, 
the $V/V_{max}$ statistics leads to constraints on cosmological parameters.
However these results rely on an arbitrary model of PLE, since the evolutionary 
effects dominate the cosmological effects in the QSO distribution, and we 
have to be sure of the reliability of the evolution model before constraining 
the cosmological parameters. So, the analysis in Van Waebeke et al. (1996) is 
rather a test of compatibility between an evolution model and a couple of 
cosmological parameters
and there is a need for an {\it independent} better understanding of
the QSO evolution before applying such cosmological tests.

A number of projects aim at assembling complete and homogeneous
samples of several thousands of quasars which will allow more reliable
and more subtle analyses, leading, thus, to verifications or 
improvement of complex theories on quasar evolution. 

Finally, there is some hope for progress in the understanding of 
the evolution of the stellar formation rate and of quasars, and in the 
determination of cosmological parameters compatible with the observed quasar 
distribution. Note that this is a necessary framework in order to make a
precise estimate of the effects of the re-ionization in the future 
temperature maps of Planck-Surveyor.

The aim of this paper is to explore in details the possibility for 
the evolution to depend on the redshift, and on whether we can distinguish 
between artefact and physical evolution. For the present study we decided 
to adopt a cosmology and an evolution model with a single evolution parameter 
which is allowed to depend on the redshift (and/or the magnitude in Section 4).
This should lead to more complex evolution laws, which may help in theoretical
understanding of quasar evolution and in turn could be the basis of
future, more realistic cosmological tests.

The technical basis of this work is the $\langle V/V_{max}\rangle $ 
test, performed in bins of
redshift. In \S \ref{samples} and \ref{evomodels}
we make a brief description of the quasar samples and the evolution models 
used. In \S \ref{evomag} we give a justification of our choice 
to study quasar evolution in redshift bins and in \S \ref{m1} we present the 
different redshift binning modes used for our study.
\S \ref{simulcat} contains the results obtained from the simulated catalogues. 
\S \ref{kvsz} contains the results for the real quasar
samples. More precisely, it shows the measured dependence of the 
evolution parameter versus redshift, for a few sets of cosmological parameters. 
Finally, in \S \ref{discuss} we discuss the main results of our study. 

\section{Samples used}
\label{samples}

      In this study two complete quasar samples have been
used, the AAT catalogue of Boyle et al. (\cite{Ba}) and the Large Bright
Quasar Survey catalogue (\cite{Ha}), hereafter AAT and LBQS respectively.
The AAT catalogue contains 400 faint quasars selected by their
UV excess, it is claimed to be complete within $18 \le m \le 21$ 
and $0.2\le z \le 2.29$.
LBQS consists of 1055 optically selected quasars of intermediate
brightness, with $16.5 \le m_{B_J} \le 19$, within the redshift
range $0.2\le z \le 3.4$. Various methods have been used for the
selection of this catalogue, combining a color selection
technique and slitless spectroscopic data. LBQS is almost
three times bigger in size and much more profound in redshift than
AAT, but its selection criteria depend on the redshift. This
means that the selection biases differ from one bin of redshift to another,
which will lead to dramatic effects on the measured evolution.
In the case of AAT, the UVX selection technique applied in the whole
redshift range insures the fact that the biases can be comparable throughout
the whole sample. The differences between the two samples may
influence the results on quasar evolution
thus a direct comparison may not be possible.

\section{Evolution models and evolution parameter}
\label{evomodels}

\subsection{PLE and PDE}
In order to explain the observed non-uniform spatial distribution of
quasars, several 
empirical evolution models have been proposed such as the PLE (Pure
Luminosity Evolution) model, the PDE (Pure Density Evolution model) and 
the LDDE (Luminosity Dependent Density Evolution). In this paper, we
will make an extensive use of PLE and PDE, whose introduction was 
purely phenomenological. PLE makes the hypothesis of a constant object 
space density and supposes the following form for the luminosity:
\begin{equation}
        L(z)=L(0)e(z)
\end{equation}
where $e(z)$ is an one parameter evolution law.
On the contrary, PDE assumes a constant quasar luminosity and a 
redshift--dependent space density:
\begin{equation}
        \rho(z)=\rho(0)e(z)
\end{equation}

\subsection{Power Law and Exponential Parametrizations}

The most commonly used expressions for $e(z)$ are the power low form
\begin{equation}
        e(z)=(1+z)^k
\label{eq3}
\end{equation}
and the exponential form
\begin{equation}
        e(z)=exp(k\,\tau(z))
\label{eq7}
\end{equation}
where $\tau(z)$ is the lookback time, and $k$ is the evolution parameter
($k_L$ and $k_D$ in the cases of a PLE and a PDE, respectively).

The basis of this work is the $\langle V/V_{max}\rangle $ test, or rather the
$\langle V_e/V_a\rangle $ test (Avni and Bahcall, 1980), with all
volumes computed according to Mathez et al., 1996.

In the following 
we examine the variations of the evolution parameter
in magnitude and redshift bins for PLE and PDE models.
All $k_L$ ($k_D$) calculated here are the optimum values
which insure a 
$\langle V/V_{max}\rangle = 1/2$ ($\langle W/W_{max}\rangle = 1/2$) 
\footnote{$W(z)$ is defined as: $dW(z)=\rho(z)dV(z)$}
within each bin, a necessary (but not sufficient) condition 
for a uniform $V/V_{max}$ ($W/W_{max}$) distribution within the interval
$[0,1]$. The dependence of $k_L$'s ($k_D$'s) versus redshift is 
measured and all calculations are made for 
several cosmological models. During our work, we examined cosmological
models with $\Omega_0$ and $\Lambda$ varying between 0 and 1 with steps of
0.25. In the present paper we chose to demonstrate our results for the
flat univers cosmological models ($\Omega_0 + \Lambda = 1$), excluding 
the rather non--realistic model $(\Omega_0,\Lambda)=(0,1)$. In all plots
representing $k_L$ or $k_D$ versus $z$, the values of the evolution
parameter increase with increasing $\Omega_0$ and with decreasing $\Lambda$,
unless specified otherwise. So the lower curve corresponds to the model
$(\Omega_0,\Lambda)=(0.25,0.75)$ while the upper one to the model
$(\Omega_0,\Lambda)=(1.0,0.0)$. The above method allows a direct
determination of the evolution parameter error bars; the $1\sigma$ region
of $\langle V/V_{max}\rangle$,calculated assuming no evolution,  is given by
$1/2 \pm 1\sigma$ with $\sigma = (\sqrt{12N})^{-1}$, where $N$ is the
number of quasars within each bin. A corresponding error bar on
the evolution parameter can thus be derived from this $1\sigma$ interval of
$\langle V/V_{max}\rangle$. We will go back to this point in the
Appendix.

Our method differs significantly from that used by Boyle et al., 1988 on the
AAT sample. They showed that
PLE power law with $k_L \simeq 3.5$ is a good fit to the AAT sample,
with a technic which consists in calculating the
luminosity function within redshift bins and in fitting the results 
to a luminosity function model. This method has the drawback to 
consider the evolution parameter
as a free adjustable parameter, as well as the slopes of the 
luminosity function, the density,
and the absolute magnitude linking the two slopes.
It does not take into account the physical information
that the quasars should be uniformly distributed, once the evolution has been
corrected. Indeed, this information allows the evolution parameter to
be determined directly from a
$V/V_{max}$ test without any hypothesis on the luminosity function.

\section{Evolution in bins of magnitude}
\label{evomag}

\subsection{Apparent magnitude}

\begin{figure}
\centerline{
\psfig{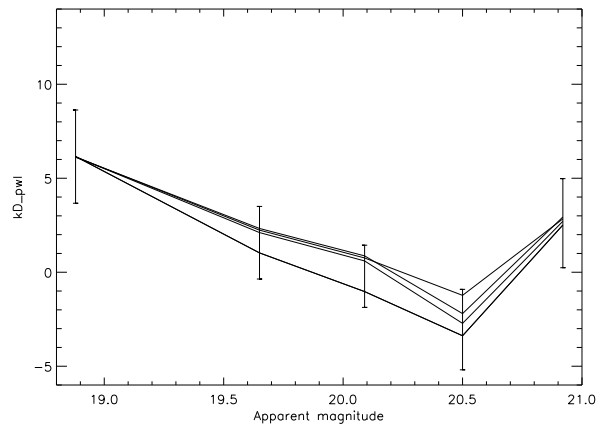}}
\caption{Variations of the evolution parameter $k_D$ with apparent
magnitude for flat universe cosmologies, with a power law evolution
model. The different curves correspond to the different cosmological 
models specified in the text.}
\label{appmag}
\end{figure}

In case of PLE it is very difficult to divide a sample in bins of
absolute magnitude, even through complex methods, because absolute magnitude 
has to be estimated at a given epoch and thus depends on the evolution law. 
On the contrary, a division in bins of apparent magnitude, $m$, is very easy 
to operate. The AAT sample 
has been divided into 5 bins which contain equal numbers of quasars. In Fig. 
\ref{appmag} we present the evolution parameter 
$k_D$ versus $m$ under a power law hypothesis. We notice a 
diminishing trend up to $m\simeq 20.5$ reversing towards fainter magnitudes. 
This trend can be a priori understood in terms of variations of 
evolution with either redshift or absolute magnitude since both are 
more or less correlated with apparent magnitude.

\subsection{Absolute magnitude}

\begin{figure*}
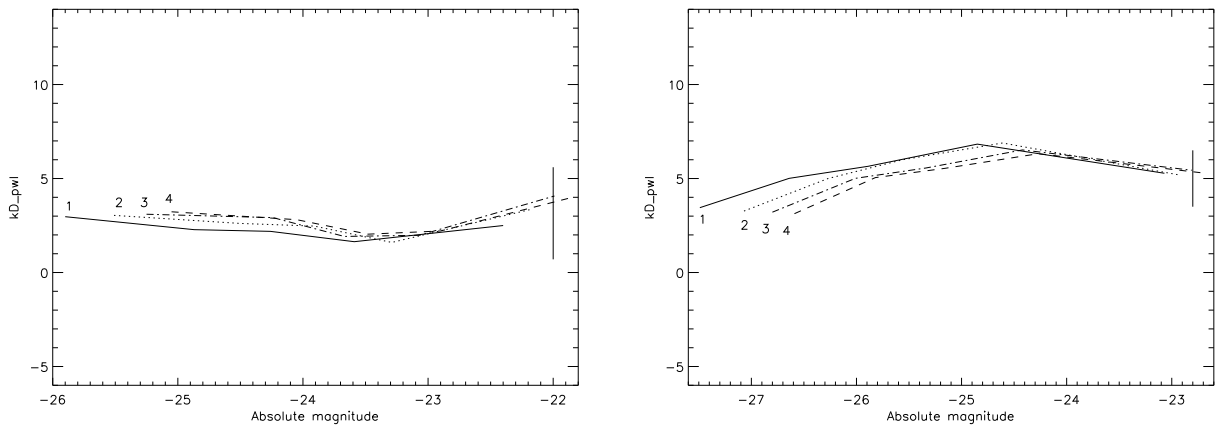

\centerline{
\psfig{figure=qso.ps2a,height=6cm,angle=0}
\psfig{figure=qso.ps2b,height=6cm,angle=0}
}
\caption{LDDE for a power law evolution model, for AAT (left) and LBQS
(right): variations of the evolution
parameter $k_D$ with absolute magnitude for 4 flat universe cosmological 
models: 1) ($\Omega_0,\Lambda$)=(0.25,0.75), 2) ($\Omega_0,\Lambda$)=(0.5,0.5), 
3) ($\Omega_0,\Lambda$)=(0.75,0.25) and 4)
($\Omega_0,\Lambda$)=(1.0,0.0). There is no clear trend for $k_D$ with
absolute magnitude. The bar on the right gives the error bar's size ($1
\sigma$.}
\label{mag}
\end{figure*}

Analyzing a sample of
quasars with a PDE hypothesis, and with $M_0<-23$ for the cosmological models with $q_0=0.1$ and
$q_0=0.5$, Schmidt and Green (1983) derived that the evolution parameter
$k_D$ increases
with absolute magnitude. Making a similar analysis on AAT and LBQS
catalogues (which span a larger absolute magnitude range),
we find no clear tendency for the evolution parameter $k_D$ versus
absolute magnitude, whatever the cosmological model used. The samples have
both been divided in 5 bins of absolute magnitude with a constant number
of objects ($\sim 80$ for AAT, $\sim 210$ for LBQS). Fig. \ref{mag} 
illustrates the results for the zero curvature cosmological models
examined above. On the right side of the figures we present a typical 
error bar. Note that the scaling of the vertical
axis is the same in both Figs. \ref{appmag} and \ref{mag}. 

From this previous analysis, we conclude that Fig. \ref{appmag} could
reflect a redshift dependence, not an absolute magnitude effect, 
and that is why this
paper is devoted to the redshift dependence of the evolution.

\section{Bins of Redshift}
\label{m1}

The quasar samples have been divided into several redshift bins in order
to examine the redshift dependence of the evolution parameter
and the influence of the cosmological model.
The AAT analysis is limited to the redshift interval $[0.3,2.2]$,
because, at low and high redshift, the sample is
obviously not complete. The binning intervals are defined here.

The binning was chosen so as to be optimized with respect to 
different criteria. The first one is 
that bins should contain a sufficient number of quasars in order to
make a reliable statistical analysis. The second one is that they
should be narrow enough so that the evolution parameter may be considered 
as constant within a single bin. The third criterion is that,
according to the parametrization of the evolution hypothesis, 
we must chose the binning in order to distribute the amount of 
evolution equally within bins.

Our first choice of binning insures equal numbers of quasars, in order to
distribute equally the statistical noise between bins and to minimize
it, thus satisfying the first criterion above. The AAT sample was divided
in 5 bins of about 80 quasars each (binning (B1), Table \ref{T1}) while the 
LBQS sample was divided in 11 bins each one containing about 100 quasars.
Redshift limits for LBQS are: 0.20, 0.37, 0.57, 0.75, 0.96, 1.14, 1.36,
1.81, 2.10, 2.52, 3.36.

We must be sure however that the variations of the observed evolution we are 
looking for are not biased by the binning chosen. It could be the case
if the evolution rate differed too much from bin to bin.
This is why, for each assumed evolution law $e(z)$, we look for
a binning ensuring equal ratios $e(z_{i+1})/e(z_i)$.
In the case of a power law parametrization for example, used to 
describe most of the evolution laws derived 
for quasars, the redshift enters through the factor $(1+z)^{k}$ (see equation
\ref{eq3}). With such a law and binning (B1) in Table 1, the ratio 
$\left(\frac{1+z_{i+1}}{1+z_i}\right)^k$ which governs the evolution rate 
within bin $i+1$, decreases strongly from bin to bin and this is likely to bias 
the analysis. If we want to take into account the dependence of the
evolution rate with redshift, we must impose the redshift intervals'
limits, $z_i$, such as: 
\begin{equation}
\frac{1+z_{i+1}}{1+z_i}=\frac{1+z_{i+2}}{1+z_{i+1}}
\label{eq1}
\end{equation}
With this binning (B2), the evolution should be comparable within all 
bins and we should be expecting a somehow more uniform 
$\langle V/V_{max}\rangle$ distribution,
provided that $k$ does not vary too much from bin to bin.

\begin{table}
\caption{Bins' redshift limits and quasar numbers for AAT sample. The $z_i$
for the third binning have been computed for the cosmological model
$(\Omega_0,\Lambda)=(0.5,0.5)$.} 
\label{T1}
\begin{center}
\begin{tabular}{cccccccc}
\hline
(B1)&$z_i$&0.30&0.86&1.22&1.57&1.87&2.2\\
&$n$&&78&78&78&78&78\\
\\
(B2)&$z_i$&0.30&0.56&0.86&1.23&1.67&2.2\\
&$n$&&19&66&72&104&124\\
\\
(B3)&$z_i$&&0.30&0.51&0.80&1.27&2.2\\
&$n$&&&15&47&99&222\\
\hline
\end{tabular}
\end{center}
\end{table}

The exponential parametrization leads to another binning, also 
given in Table \ref{T1} (binning (B3)) for the cosmological model 
$(\Omega_0,\Lambda)=(0.5,0.5)$. Here, the look--back time intervals are
constant within bins (note that the bins' limits depend slightly on the 
cosmological model). We chose to divide the sample 
in 4 bins because of the very small numbers of quasars contained in the first 
two bins in a choice of a division in 5 redshift intervals (13 and 33). We 
remind the reader that a high number of quasars within bins is very important 
because of the statistical nature not only of the test, but of the
evolution itself, which is a further motivation
for large redshift bins. But too large a bin may bias
the evolution parameter estimation if the evolution rate changes too strongly
with the redshift, which may be the case in the last redshift interval
for this binning. 

\section{Tests with simulated catalogues}
\label{simulcat}

The method has first been validated with Monte-Carlo simulated catalogues.
The aim of these tests was to examine the results obtained by analyzing 
simulated catalogues, constructed under specific
hypotheses for the whole redshift interval, whose properties are
{\it a priori} known. Here we present two different tests
applied to simulated catalogues whose characteristics mimics the
properties of AAT. The details of the simulation algorithm can be found
in Mathez et al. 1996.

\begin{figure}
\centerline{
\psfig{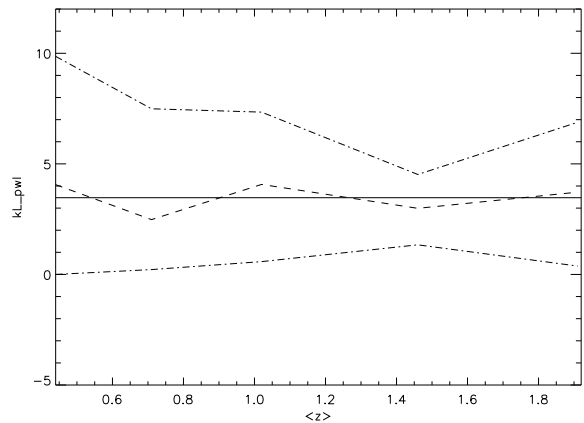}
}
\caption{$k_L$ versus $z$ for 38 simulated catalogues, binning B2. 
The dashed--dotted
line includes the 68\% of the $k_L$'s values, while the dashed line illustrates
the variations with redshift of the median value. All catalogues were
constructed under a PLE, power law hypothesis of $k_L=3.47$ for the
cosmological model $(\Omega_0,\Lambda)=(0.5,0.5)$.}
\label{simpwl}
\end{figure}

For the first test, 38 simulated catalogues of 400 quasars each, were 
built under a constant luminosity evolution hypothesis, a power law 
parametrization with $k_L=3.47$, 
within the redshift region $z \in [0.3,2.2]$. The cosmological model 
used for the construction was $(\Omega_0=0.5,\Lambda=0.5)$.
Fig. \ref{simpwl} shows the measured evolution in redshift bins (with the
binning B2) on these
catalogues. The solid line indicates the true value $k_L=3.47$, the
dashed--dotted line includes the 68\% of the
values and the dashed line corresponds to the variations of the median $k_L$
value with redshift.
The analysis with binning (B1) on the same catalogues
gives a similar result, which proves that
for a constant evolution rate, the method is not sensitive to the chosen
binning.

\begin{figure}
\centerline{
\psfig{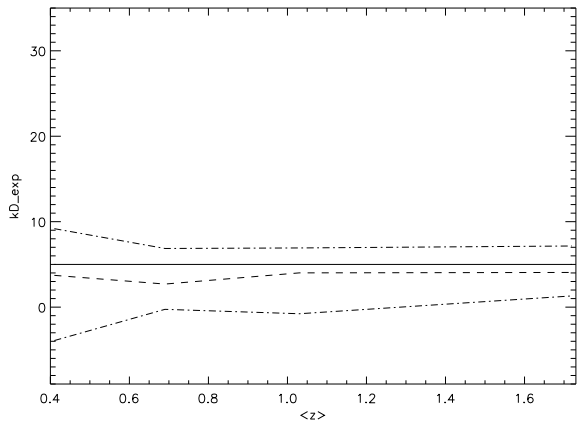}
}
\caption{$k_D$ versus $z$ for 44 simulated catalogues, binning B3. 
The dashed--dotted
line includes the 68\% of the $k_D$'s values, while the dashed line
illustrates
the variations with redshift of the median value. All catalogues were
constructed under a PLE, exponential law hypothesis of $k_L=5$ for the
cosmological model $(\Omega_0,\Lambda)=(0.5,0.5)$, and analysis is made
under PDE exponential law hypothesis.}
\label{simexp}
\end{figure}

For the second test, 44 simulated catalogues were build under a PLE,
exponential parametrization ($k_L=5$), for the same cosmological model
as above. 
The analysis has been made for binning (B3) and assuming a (wrong)
PDE hypothesis. The results on Fig. \ref{simexp} shows that a density
evolution is measured, with $k_D$ slightly lower than the true $k_L$.
It should not be surprising, indeed with a $V/V_{max}$ method, it is
impossible to separate luminosity and density evolution. On the contrary,
the distinction between these two kinds of evolution is possible with methods
based on luminosity function hypothesis (as in Boyle et al., 1988), but as
explained in Section 3, it suffers from some other problems.

\section{Evolution parameter versus $z$}
\label{kvsz}

The evolution parameter is now calculated within each redshift bin, by
forcing that $\langle V/V_{max}\rangle=0.5$ 
(or $\langle W/W_{max}\rangle=0.5$). An error bar is also
derived, from the $1\sigma$ interval of $V/V_{max}$, according to the method
described in details in the Appendix.

\subsection{PLE results}
\label{r2}

Fig. \ref{B2} shows $k_L(z)$ for AAT catalogue for power law and exponential 
forms, and zero curvature cosmological models. Binning (B1) is the first to 
be examined.

\begin{figure*}
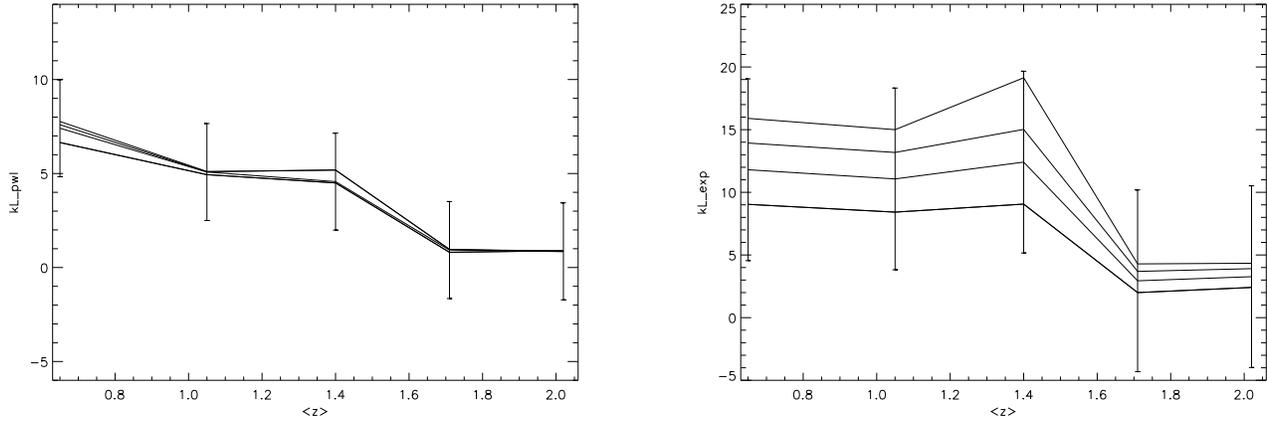

\centerline{
\psfig{figure=qso.ps5a,height=6cm,angle=0}
\hskip 0.7cm
\psfig{figure=qso.ps5b,height=6cm,angle=0}
}
\caption{AAT: The sensitivity of $k_L$ to cosmological models and its
variations with redshift, for equally populated bins (binning B1), 
for a power law parametrization (left graph) and the
exponential parametrization (right graph). The models treated are those
with zero curvature.} 
\label{B2}
\end{figure*}

\begin{figure*}
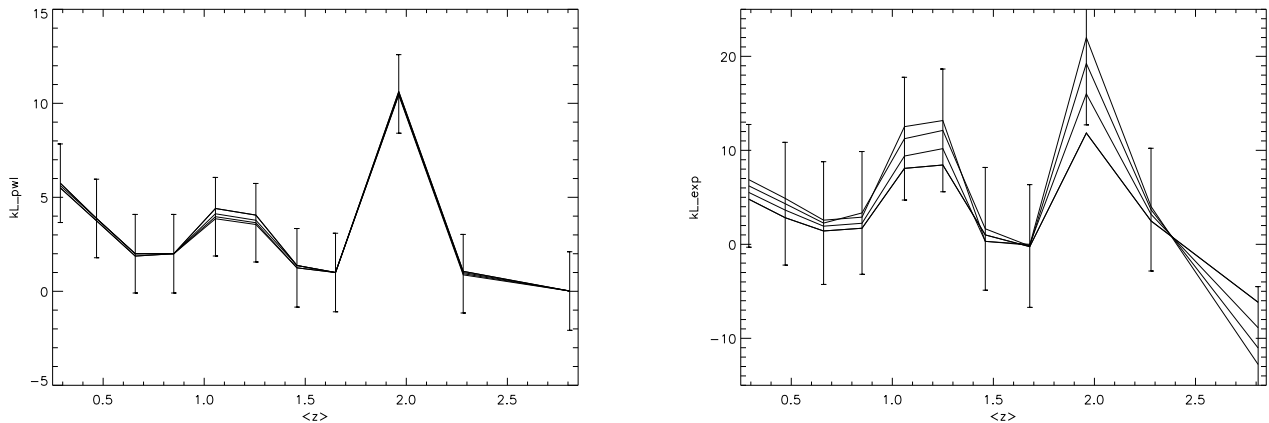

\centerline{
\psfig{figure=qso.ps6a,height=6cm,angle=0}
\hskip 0.7cm
\psfig{figure=qso.ps6b,height=6cm,angle=0}
}
\caption{The sensitivity of $k_L$ to cosmological models and its
variations with redshift for the LBQS analysis in equally populated bins
(binning B1). The left graph corresponds to a
power law hypothesis; the right one to an exponential form. The models treated 
are those with zero curvature. The fluctuations seen above are most probably 
due to the passage of strong emission lines through the filter's mean wavelength.}
\label{L2}
\end{figure*}

The left plot corresponds to
a power law and the right one to the exponential law parametrization. In 
both cases $k_L$ decreases slightly with redshift. The same analysis on the 
LBQS (Fig. \ref{L2}), results in a figure dominated by strong
fluctuations, absent in the result given by the AAT (Fig. \ref{B2}).
The fact that AAT and LBQS have not the same magnitude range is not enough to 
explain why the evolution is so different between these two catalogues.
Fig. \ref{L2} also shows a
global decreasing trend of $k_L$ with increasing redshift, which is 
marginally significant accounting for the error bars. 

The reason for this discrepancy should first be searched in the incompleteness
and/or the selection bias of the catalogues, in particular in the LBQS which has
a complex selection criteria. The combination
used (hypersensibilized IIIa-J+GG395) had a maximum width of 2000 \AA \, 
with a maximum response at
a wavelength of around 4200 \AA \, and a blue limit imposed by
the atmospheric cut-off at $\sim$ 3200 \AA \, while the red one was due to
the emulsion. Objects either with a detected
emission line or whose median wavelength of 
the SED was found bluewards of the central wavelength were classified as 
quasar candidates while the others were rejected (\cite{Ha}).
The most intense emission lines of quasars in the waveband 
we are interested in along with their relative intensities are listed
in Table \ref{T3}. 

\begin{table}
\caption{Quasars' strong emission lines and relative intensities taken
from current literature.}
\label{T3}
\begin{center}
\begin{tabular}{ccc}
\hline 
Emission&wavelength (\AA)&relative\\
\, line&\,\,\,\,\,&intensity\\
\hline 
Ly$\alpha$&1216&1\\
C{\sc iv}&1549&0.5\\
C{\sc iii}$]$&1909&0.2\\
Mg{\sc ii}&2798&0.2\\
\hline
\end{tabular}
\end{center}
\end{table}

\begin{figure}
\centerline{
\psfig{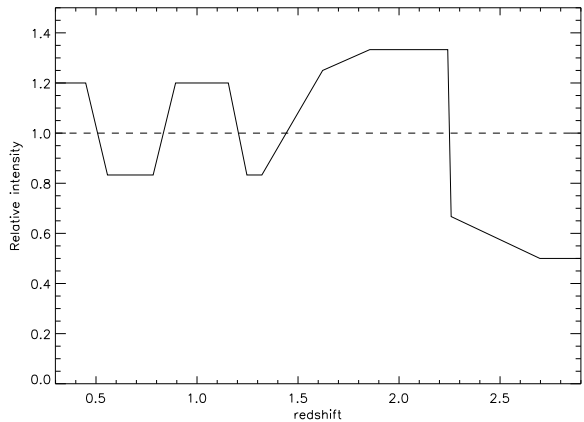} }
\caption{Relative excess and deficiency of apparent evolution due to
strongest emission lines in the LBQS.}
\label{Pr1}
\end{figure}

As redshift increases, 
the quasar strong emission lines appear successively in the blue 
and in the red part of the
LBQS spectral range, leading each time to a shift of the median
wavelength. This gives rise first to an excess and 
then to a deficiency of quasar candidates. 
Fig. \ref {Pr1} gives our estimation on this effect on the detection
rate, and finally in the evolution parameter $k_L$. 
We estimated that the filter's response gets its 
maximum value at a wavelength of $\sim 3700$ \AA \, and that it remains almost
constant up to a $\lambda \sim 4900$ \AA. Based on these hypotheses we first
calculated the redshift regions where each emission line would contribute 
to an excess (and to a deficiency) depending on its relative intensity. The
conversion of wavelengths to redshifts has been made by using the forms:
\begin{equation}
\label{eq4}
(\lambda_L + \Delta\lambda_L)(1+z_{1L})=\lambda_1
\end{equation}

\begin{equation}
\label{eq5}
(\lambda_L \pm \Delta\lambda_L)(1+z_{2L\pm})=\lambda_2
\end{equation}

\begin{equation}
\label{eq6}
(\lambda_L -\Delta\lambda_L)(1+z_{3L})=\lambda_3
\end{equation}

with $\lambda_1=3700$ \AA \,, $\lambda_2=4200$ \AA \, 
and $\lambda_3=4900$ \AA\,, $\lambda_L$ the wavelengths given in Table \ref{T3}
and $\Delta\lambda_L$ the Full Width Half Maximum (FWHM) of each line.
$z_{1L}$ is the redshift at which 
the emission line $L$ enters the detection range, contributing to an apparent 
quasar excess. $z_{2L+}$ marks the end of this contribution. At $z=z_{2L-}$ the
emission line enters the filter's red region and causes an apparent
deficiency of objects and after $z_{3L}$ it is no longer detectable. 
As FWHM and EW vary from one quasar to the other and from one emission line to
the other, approximate values have been used, taken from \cite{Ul}. The
composition of Fig. \ref{Pr1} (whose similarity with Fig. \ref{L2} is
obvious), has been made by superimposing the results for the 4 emission lines.
The $k_L$'s maxima approximately coincide with the probable high detection 
rate points giving a possible explanation for the $k_L$'s fluctuations with
mean redshift. According to Hewett et al. (1995), other factors are
likely to influence also the
quasars' detection rate but their analysis 
is outside the aims of this paper. 

Right afterwards we test binnings (B2) and (B3) for AAT sample.
Fig. \ref{B4} shows the variations of $k_L$ with redshift $z$ for these two
cases. The values of $k_L$ are almost constant within error bars
in the redshift range $0.7 \le z \le 1.7$. However, the values of $k_L$ in
small redshift bins are not very reliable, as we 
can see from the first error bar
in both graphs, due to the very small number of quasars (see Table \ref{T1}).  
We repeated the procedure described above but this time we divided the sample
into 4 bins, so as to have more objects (32) in the first bin. We found that the
evolution parameter was significantly higher in this bin and constant in the
others but we cannot use these results: because of the large size of the three
other redshift intervals, the test looses its sensitivity. 
From the above we conclude that a PLE (or a PDE as shown later on), with
$k_L \simeq 3.5$ for a power law and $k_L\simeq 9$ for an exponential,
fits the sample, as already demonstrated by Boyle et al. 
(1988) (but our method differs significantly, as explained in Section 3.2).
We also applied the analysis to LBQS for the equivalent of binnings 
(B2) and (B3) as well
as for random binnings, even though we consider this sample as highly
biased. We find in average no significant decrease of $k$ with increasing 
$z$, up to $z \simeq 2$. Fig. \ref{anotherLBQS} illustrates these
results for binning (B2). Peaks are far less intense but excess and
deficiency are found in the predicted places (compare with Fig. \ref{Pr1}).

\begin{figure*}
\centerline{
\psfig{figure=qso.ps8a,height=6cm,angle=0}
\hskip 0.7cm
\psfig{figure=qso.ps8b,height=6cm,angle=0}
}
\caption{$k_L$ variations with $<z>$ for AAT, for flat universe models and 
for binning B2 given by equation \ref{eq1} (left graph) and for binning
B3 ensuring 
equal look--back time intervals (right graph). (Note the difference of the
scales of the vertical axis on the two graphs.)} 
\label{B4}
\end{figure*}

\begin{figure}
\centerline{
\psfig{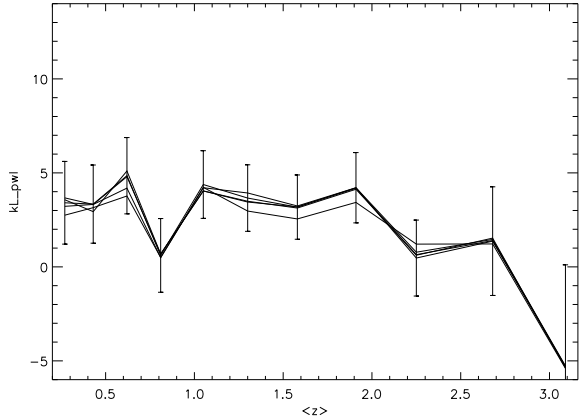}
}
\caption{$k_L$ variations with redshift for LBQS, for binning (B2).
$k_L$ is rather constant within the redshift range $0.3 \leq z \leq
2$.}
\label{anotherLBQS}
\end{figure}

\subsection{PDE results}

\begin{figure*}
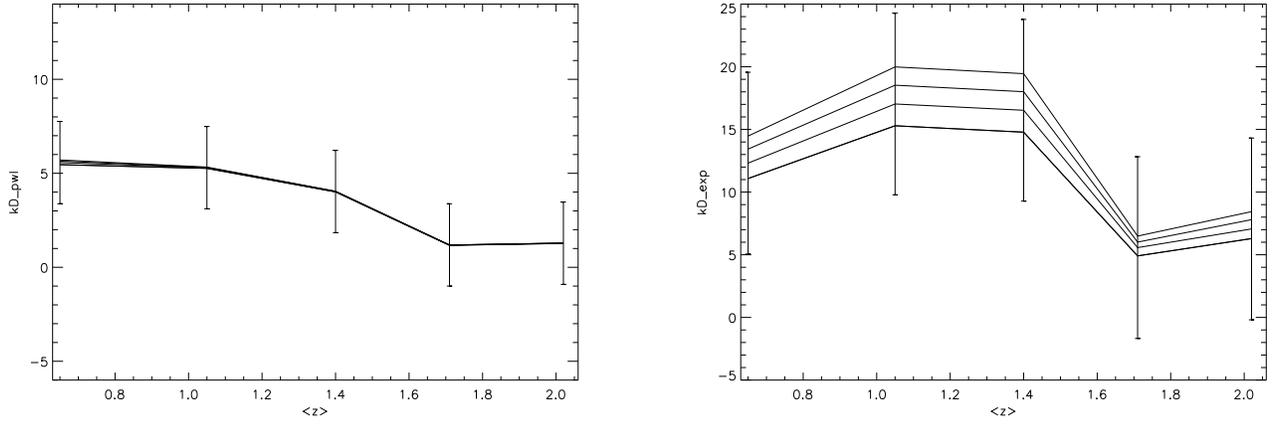

\centerline{
\psfig{figure=qso.ps10a,height=6cm,angle=0}
\hskip 0.7cm
\psfig{figure=qso.ps10b,height=6cm,angle=0}
}
\caption{$k_D$ versus $<z>$ for AAT catalogue, binning B1. 
Left graph: power law PDE, right graph: exponential PDE.} 
\label{B5}
\end{figure*}

\begin{figure*}
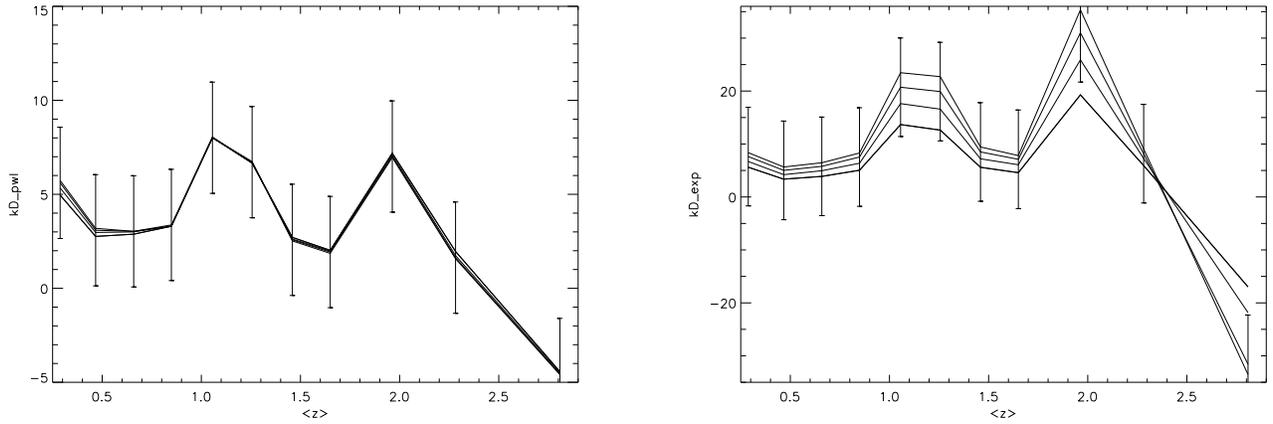

\centerline{
\psfig{figure=qso.ps11a,height=6cm,angle=0}
\hskip 0.7cm
\psfig{figure=qso.ps11b,height=6cm,angle=0}
}
\caption{$k_D$ versus $<z>$ for LBQS catalogue, binning B1. 
The left (right) graph corresponds to the power law (exponential) 
parametrization.}
\label{L3}
\end{figure*}

\begin{figure*}
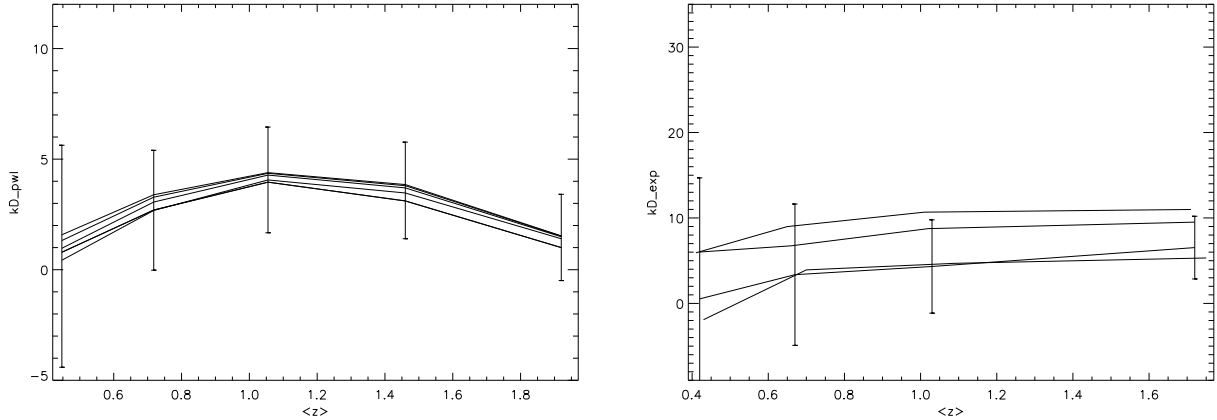

\centerline{
\psfig{figure=qso.ps12a,height=6cm,angle=0}
\psfig{figure=qso.ps12b,height=6cm,angle=0}
}
\caption{The analysis of AAT catalogue: $k_D$ versus $<z>$ for flat universe 
models and for binnings (B2) and (B3) given in Table \ref{T1}, for the flat
universe models.}
\label{B6}
\end{figure*}

The same work has been made under the hypothesis of a PDE, 
for both power law and exponential laws. In Figs. \ref{B5}
and \ref{B6} we present the results for AAT for the three binnings
and in Fig. \ref{L3} the results for LBQS for the first binning.

In all cases we find the same characteristics and the same trends as 
for the PLE hypothesis. As already mentioned in Section 6 about 
the simulated catalogues, a density evolution almost as strong as a 
luminosity evolution is measured because of the degeneracy between 
these two kinds of evolution, which
is not broken by the $V/V_{max}$ method.
All figures concerning binning (B1) for AAT point out 
that whatever the hypothesis on the evolution law, the evolution parameter 
declines slightly with redshift. The 
similarity between the results of the PLE and PDE hypotheses indicates that 
both evolution models fit equally AAT sample. 
In the case of LBQS the evolution
parameter's peaks become less intense but they are always present.

Furthermore, the small
variations with redshift of $k_L$ and $k_D$ imply that both models, 
when assuming a constant evolution parameter, are not far from being
correct, at least in the redshift range $[0.7,1.7]$. With binning (B1)
for AAT however, the picture (Fig. \ref{B2} and \ref{B5}) 
was rather of a regular decrease of the evolution parameters 
with increasing $z$. In order to check this point we
made the same analysis in a series of random binnings, for both AAT and
LBQS. We found that most binnings
rather point to a constant evolution parameter, at least up to  $z=1.7$
and to a small tendency towards a regular decrease above this redshift. 

\subsection{Summary}

\begin{figure}
\centerline{
\psfig{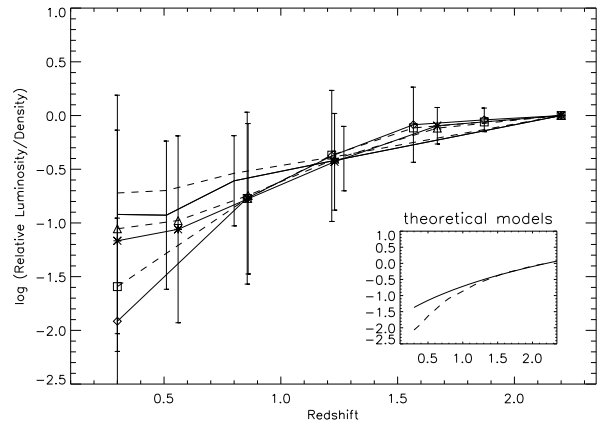}
}
\caption{Relative Luminosity (solid lines)/density (dashed lines) normalized to
$z=2.2$ versus $z$. $\Box$,$\Diamond$: (B1), power law parametrization. 
$\triangle$,$\ast$: (B2), same parametrization. Lines without plotting symbols:
(B3), exponential parametrization. Inset figure: the solid line
represents a power law parametrization of $k=3.5$ while the dashed 
one represents an exponential parametrization of $k=9$.}
\label{rellum}
\end{figure}

We will now give a different representation of our results concerning AAT, by
calculating the relative luminosity and density, normalized at $z=2.2$, for
both PLE and PDE models. Fig. \ref{rellum} 
illustrates $log \,L(z)/L(2.2)$ (solid lines) and $log \, \rho(z)/\rho(2.2)$
(dashed lines) versus $z$. Diamonds and squares correspond to binning (B1) 
given in Table \ref{T1}, for a power law parametrization. Triangles and stars 
correspond to binning (B2), for the same parametrization. The two curves
without symbols represent the third binning, (B3), and an exponential
parametrization. Error bars have been plotted for each binning. 
As it may be noticed, they become wider as the redshift is smaller because 
of the
propagation of errors in the calculation of the normalization constants.
All calculations have been made for the $(\Omega_0,\Lambda)=(0.5,0.5)$ model.
There is a remarkable similarity with previous results 
(see e.g. Fig. 7 of Shaver (1994), Fig. 3 of Pei (1995) and
Fig. 1 of \cite{Sa98}, all obtained from different
procedures) as well as with the theoretical results of Yi (1996, Fig. 2).
In the inset figure we present the same quantities calculated for a
constant evolution parameter. The solid line represents a power law
parametrization of $k=3.5$ while the dashed line represents an
exponential parametrization of $k=9$.

\section{Discussion}
\label{discuss}

This paper presents an analysis of the evolution rate of
quasars at various epochs. Our method differs from most previous ones in
that it does not make use of any determination of luminosity function.
It solely consists in computing
the evolution parameters $k_L$ and $k_D$ (assuming PLE and PDE respectively)
in bins of redshift, such that the $V/V_{max}$ are uniformly
distributed. However, the various selection criteria applied in
the construction of the two catalogues (AAT and LBQS) as well as the different
redshift and magnitude limits make a direct comparison of the results
quiet difficult. The method has been tested with Monte-Carlo simulated
catalogues.
Three binnings have been adopted, the first one with equal numbers per bin,
the second and the third with a priori similar evolution rates inside
each bin, according to the evolution law.
Our results are the following:

\,\, $\bullet$ Both samples roughly show the same large trends, however 
modulations do
appear in the LBQS results, which are likely to be correlated with the
crossing of the main emission lines from the blue to the red side of the 
available spectral range.

\,\, $\bullet$ All results on the evolution parameters 
are quite similar whatever the hypothesis, PLE or PDE, 
mainly due to the inefficiency of the 
$V/V_{max}$ test to distinguishing between density and luminosity
evolution. 

\,\, $\bullet$ Similar results are also obtained for both power law and
exponential parametrizations. Certainly a single phenomenon cannot be
described by two different laws but we note that there is no significant
difference between these laws in the redshift range $z \in [0.7,2.2]$,
as seen in the inset plot in Fig. \ref{rellum}. The determination of
quasars' evolution rate at high redshifts will also determine which of
the proposed models is the appropriate one (if any).

\,\, $\bullet$ In bins with equally distributed evolution, however,
both parameters $k_L$ and $k_D$ show far less variations with redshift
and within error bars we can suppose that they are constant ($2.5 \leq
k_L \leq 4, 1.5 \leq k_D \leq 3$ in a power law parametrization). 

\,\, $\bullet$ $k_L$ ($k_D$) and $\langle V/V_{max}\rangle$ are
linearly correlated, as shown in Fig. \ref{correl1}.

We confirm that a PLE (or PDE) with a constant evolution parameter is a
good approximation for the redshift range $z \in [0.7,1.7]$. 
At larger redshift, the results of our analysis of the LBQS are
consistent with a maximum luminosity (or density) around $z=2.5$, but the
reliability of these results is not yet established due to likely selection
biases. The determination 
of this $k(z)$ dependence towards larger $z$ is essential for the 
understanding of the birth and growth of quasars at
high redshifts, and the relation of the quasar phenomenon with star
bursts in the primordial universe. 

\section*{Appendix}
\begin{figure}
\centerline{
\psfig{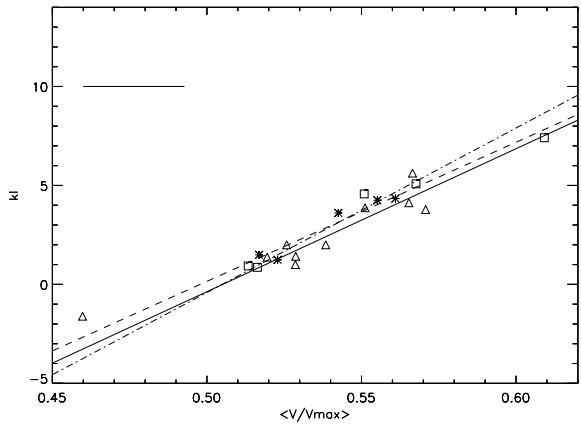}
}
\caption{$k_L$ versus $\langle V/V_{max}\rangle$ in the case of a 
power law parametrization.
- - -, $\Box$: AAT's first binning. -$\cdot$-, $\ast$: AAT's second binning. 
---, $\triangle$: LBQS. All calculations have been made for the cosmological 
model $(\Omega_0,\Lambda)=(0.5,0.5)$. In the upper-left corner,
a typical error bar $\sigma_{\langle V/V_{max}\rangle}= 1/\sqrt{12N} $.} 
\label{correl1}
\end{figure}

A rough estimation of the errors on $k_L$ and $k_D$
has been made, using the correlation we found
between $\langle V/V_{max}\rangle$ and $k_L$ or $k_D$
values. $\langle V/V_{max}\rangle$ has been calculated for each bin
making the hypothesis of a zero evolution. $k_L$ versus 
$\langle V/V_{max}\rangle$ is shown in Fig. \ref{correl1}.
The dashed line and the squares
correspond to binning (B1) of AAT catalogue. The dashed-dotted line and
the stars correspond to AAT's binning (B2). The solid line and the 
triangles are the LBQS results. On the upper left we show a typical error
bar on the $\langle V/V_{max}\rangle$ estimated as $\sigma_{\langle
V/V_{max}\rangle}=(\sqrt{12N})^{-1}$. 

The coefficients of the linear approximation $k_{L_{pwl}}=\alpha_L+\beta_L 
\,\langle V/V_{max}\rangle$ are given in Table \ref{T4} (also for $k_D$),
for AAT (binnings (B1) and (B2)) and LBQS. We notice that there a unique 
correlation may be adopted between $k_{L_{pwl}}$ and 
$\langle V/V_{max}\rangle$, as the error bars are much more important 
than the variations of the coefficients $\alpha_L$ and $\beta_L$. 
Fig. \ref{correl1} clearly illustrates that if the quasar population
does not evolve ($k_L$ or $k_D=0$), then their $\langle V/V_{max}\rangle$
is equal to 0.5, as expected.

\begin{table}
\caption{Numerical values for the coefficients in the linear
approximation $k_{L_{pwl}}=\alpha_L+\beta_L \,\langle V/V_{max}\rangle$ and 
$k_{D_{pwl}}=\alpha_D+\beta_D \,\langle V/V_{max}\rangle$ in a PLE and 
a PDE hypothesis, respectively.}
\label{T4}
\begin{center}
\begin{tabular}{ccccc}
\hline
Catalogue/&$\alpha_L$&$\beta_L$&$\alpha_D$&$\beta_D$\\
binning\\
\hline
AAT (B1)&-35&70.29&-22.95&47.75\\
AAT (B2)&-41.97&83.12&-36.74&73.08\\
LBQS&-36.6&72.26&-51.67&102.59\\
\hline
\end{tabular}
\end{center}
\end{table}

In the case of an exponential parametrization and in order to define the 
linear correlation, we must take under consideration the differentiation 
of the evolution parameters with cosmological models.

The existence of a correlation between these two quantities was somehow
expected, in the sense that the higher the value of $\langle
V/V_{max}\rangle$ is under a zero evolution hypothesis, the higher 
$k_L$ (or $k_D$) must be in order to insure a $\langle V/V_{max}\rangle
= 0.5$ in a PLE (or PDE) hypothesis.

\end{document}